\documentclass[aps,twocolumn,
showpacs,pra]{revtex4}
\usepackage{amsmath,latexsym,amssymb}
\usepackage{amsmath}
\usepackage{latexsym}
\usepackage{amssymb}

\begin{document}
\title{Multipartite positive-partial-transpose inequalities 
exponentially stronger than local reality inequalities}
\author{Koji Nagata}
\affiliation{
Department of Physics, Korea Advanced Institute of Science and Technology,
Daejeon 305-701, Korea
}

\pacs{03.65.Ud, 03.67.Mn, 03.67.-a}
\date{\today}

\begin{abstract}
We show that positivity of {\it every} partial transpose of $N$-partite quantum states implies new inequalities on Bell correlations which are stronger than standard Bell inequalities by a factor of $2^{(N-1)/2}$.
A violation of the inequality implies the system is in a 
bipartite distillable entangled state.
It turns out that a family of $N$-qubit bound entangled states proposed by 
D\"ur {[Phys. Rev. Lett. {\bf 87}, 230402 (2001)]} violates the inequality
for $N\geq 4$. 
\end{abstract}

\maketitle

\section{Introduction}

The striking feature of quantum mechanics is 
the existence of entangled states.
There is much research of nature of entangled 
states related to local realistic theories \cite{BELL,bib:Redhead,bib:Peres}.
Separable state is defined by Werner in 1989 \cite{WERNER1}.
And what state is separable or entangled was discussed very 
much \cite{DUR1,DUR2,KRAUS,KARNAS}.

In 1996, Peres-Horodecki opened the method how to classify a state in question.
A multipartite state $\rho$ has positive partial 
transposes with respect to all subsystems 
if $\rho$ is separable \cite{PPT}.
However it was shown \cite{HORODECKI} 
that this criterion is not sufficient to be separable in general.
Such an entangled state was called as bound entangled states.

It was shown that all quantum states
with positive partial transposes with respect to all
subsystems satisfy Bell inequality 
with an arbitrary number of settings per site \cite{NAGATA1}.
This result further supports the conjecture by Peres,
that all such states can admit local realistic theories \cite{PERES}.

The enormous research of quantum information theory \cite{Nielsen, Galindo}
relies on utilizing entangled pure state.
Therefore, it was discussed how to distill \cite{Bennett1,Bennett2} 
entangled pure states from a set of many entangled states with the presence of experimental noise, i.e., from a set of entangled mixed states.
In the bipartite case, one cannot distill bipartite entangled pure state 
from a set of bipartite bound entangled states because only local operations and classical communication (LOCC) cannot change the property of positivity of partial transpose which a quantum state in question has.

In the multipartite case, the situation becomes complicated.
In 2001, D\"ur proposed a family of $N$-qubit bound entangled 
states \cite{DUR}.
It was shown that 
those states violate Bell-Mermin inequality \cite{Mermin} for $N\geq 8$.
D\"ur took the definition of the bound entangled states, 
as the ones from which one cannot distill a single copy of 
some entangled pure state from a set of 
multipartite bound entangled states.
However, Ac\'\i n showed that a violation of the Bell inequality reveals bipartite distillable entanglement \cite{ACIN}. 
That is, there is at least one bipartite splitting of the system such that 
the state becomes distillable, i.e., 
one can distill a single copy of 
bipartite entangled pure state from a set of 
multipartite bound entangled states.
Therefore, D\"ur bound entangled states are bipartite distillable entangled states for $N\geq 8$.
After that, it was shown that 
D\"ur bound entangled states violate Bell inequality \cite{ZUK_KASZ} with three settings per site for $N\geq 7$ \cite{KASZLI}.
Just after that, it turned out that D\"ur bound entangled states violate Bell inequality \cite{Zukowski0} with a continuous range of settings of the apparatus at each site for $N\geq 6$ \cite{ADITI}.

And recently, it was shown \cite{ROY} that separability of $N$-partite quantum states implies new inequalities on Bell correlations which are stronger than standard Bell 
inequalities \cite{Mermin,Zukowski1,Werner2} by a factor of $2^{(N-1)/2}$.
So far, the relation between 
positivity of partial transpose of quantum states and 
Bell-Mermin inequality 
was researched by Werner and Wolf \cite{Werner3}.
However, it has not yet reported our knowledge between 
the optimal upper bound of Bell-Mermin inequality and positivity of partial transpose of quantum states in detail.

In this paper, 
we show that positivity of {\it every} partial transpose 
of $N$-qubit quantum states 
implies new inequalities on Bell correlations 
which are stronger than 
standard Bell inequalities by a factor of $2^{(N-1)/2}$.
We may call it `positive partial transpose inequalities'.
One can also see that 
a violation of the new inequality which is tested by many parties
implies that the system is in bipartite distillable entangled state
since there is at least one bipartite splitting of the system such that 
the state becomes distillable as discussed in Ref.~\cite{ACIN}.
It turns out that D\"ur bound entangled states 
violate the new inequality for $N\geq 4$.
Thus, a family of D\"ur bound entangled states has a property 
that they are bipartite distillable entangled states for $N\geq 4$.

This paper is organized as follows.
In Sec.~\ref{PPT_INEQ}, we derive a positive partial transpose inequality.
And we shall show the new inequality is stronger than standard Bell inequalities by a factor of $2^{(N-1)/2}$.
In Sec.~\ref{BOUND}, it will be shown that D\"ur bound entangled states 
violate the new inequality for $N\geq 4$.
One will see that a family of D\"ur bound entangled states has a property 
that they are bipartite distillable entangled states for $N\geq 4$.
A short summary follows
in Sec.~\ref{summary}.


\section{Positive Partial Transpose inequality}\label{PPT_INEQ}

In this section, we briefly review partial transpositions,
next present a positive partial transpose inequality.
If the partial transposes 
with respect to all subsystems are positive, the inequality is satisfied. 
And finally we show that positivity of {\it every} partial transpose 
of $N$-qubit quantum states 
implies new inequalities on Bell correlations 
which are stronger than 
standard Bell inequalities by a factor of $2^{(N-1)/2}$.

The partial transpose of an operator on a 
Hilbert space ${H}_1\otimes{H}_2$ is defined by:
\begin{eqnarray}
\left(\sum_l A^1_l\otimes A^2_l\right)^{T_1}=\sum_l{A_l^1}^{T} \otimes A_l^2,
\end{eqnarray}
where the superscript $T$ denotes transposition in the given basis.
The positivity of partial transpose is 
found to be a necessary condition for separability \cite{PPT,HORODECKI}. 
The operator obtained by the partial transpose of any separable state
is positive (PPT - positive partial transpose).
In the bipartite case of two qubits
or qubit-qutrit system,
the PPT criterion is also sufficient for separability.

In the multipartite case the situation complicates
as one can have many different partitions into set of particles,
for example four particle system $1234$ can be split, e.g., 
into $12-34$ or $1-2-34$.
Suppose one splits $N$ particles into $p$ groups,
take as an example the split into three groups $1-2-34$.
The state is called $p$-PPT
if it has positive \emph{all possible} partial transposes.
Fortunately,
positivity of partial transpose 
with respect to certain set of subsystems
is the same as positivity 
with the respect to all remaining subsystems.
In the example one should check the positivity
of operator obtained after transposition of
subsystem $1$, next $2$, and finally $34$.

In what follows, we derive positive partial transpose inequality utilising 
Bell-Mermin operator \cite{Werner3}.
All the $p$-PPT states were recently shown 
to satisfy the following inequalities \cite{NAGATA2}:
\begin{equation}
{\rm Tr}\left[ \left( |\psi^{\pm} \rangle \langle \psi^{\pm}| 
- (1-2^{2-p}) |\psi^{\mp} \rangle \langle \psi^{\mp}| \right) \rho \right] \le 2^{1-p}.
\end{equation}
If $|\psi^+ \rangle$ appears in the first term within the trace,
$|\psi^- \rangle$ appears in the second term, and vice versa.
Here, $|\psi^{\pm } \rangle$ is the Greenberger, Horne, and Zeilinger 
(GHZ) state \cite{GHZ}:
\begin{equation}
|\psi^{\pm } \rangle = \frac{1}{\sqrt{2}}
\Big[ |0\rangle_1 ... |0\rangle_N \pm  |1\rangle_1 ... |1\rangle_N \Big].
\end{equation}
Omitting the positive factor 
$2^{2-p} {\rm Tr}\left( |\psi^{\mp} \rangle \langle \psi^{\mp}| \rho \right)$
one arrives at another operator form as follows:
\begin{equation}
\Big| {\rm Tr}\left[ \left( |\psi^+ \rangle \langle \psi^+| 
- |\psi^- \rangle \langle \psi^-| \right) \rho \right] \Big| \le 2^{1-p}.
\end{equation}
Let us put $p=N$. One has
\begin{eqnarray}
\Big| {\rm Tr}\left[ \left( |\psi^+ \rangle \langle \psi^+| 
- |\psi^- \rangle \langle \psi^-| \right) \rho \right] \Big| \le 2^{1-N}\label{PPT}.
\end{eqnarray}
Of course, every state which has positive partial 
transposes with respect to all 
subsystems satisfies the above condition (\ref{PPT}).
Using the form of the Bell-Mermin operator
with two orthogonal settings par site \cite{NAGATA1}
\begin{eqnarray}
B_N=2^{(N-1)/2}(|\psi^+ \rangle \langle \psi^+| 
- |\psi^- \rangle \langle \psi^-| ),
\end{eqnarray}
the upper bound of the Bell-Mermin inequality,
for $N$-PPT states, is found to read:
\begin{equation}
|{\rm Tr}(B_N \rho^{\rm PPT})| \leq \frac{1}{2^{(N-1)/2}}
\label{NPPT_BOUND}
\end{equation}
and it can never reach the local realistic bound for $N\geq 2$.
Clearly, the inequality (\ref{NPPT_BOUND}) is 
positive partial transpose inequality.
A violation of the inequality 
implies that the system is in bipartite distillable entanglement 
since there is at least one bipartite splitting of the system such that 
the state becomes distillable \cite{ACIN}.

We shall show the inequlity (\ref{NPPT_BOUND}) is stronger than 
standard Bell inequalities by a factor of $2^{(N-1)/2}$.
Suppose the following state 
\begin{eqnarray}
\rho_V=
V|\psi^+ \rangle \langle \psi^+| +(1-V)\rho_{\rm noise}~(0\leq V\leq 1).\label{GHZ_WERNER}
\end{eqnarray}
$\rho_{\rm noise} = \frac{1}{2^N} \openone$ is the random noise admixture. The value of $V$ can be interpreted as the reduction factor of the interferometric contrast observed in the multi-particle correlation experiment.
It was shown that if the following condition 
\begin{eqnarray}
\sum_{i_1,\ldots,i_N=1}^2\left({\rm Tr}[\rho\sigma_{i_1}\otimes
\sigma_{i_2}\otimes \cdots \otimes\sigma_{i_N}]\right)^2\leq 1\label{Z}
\end{eqnarray}
is sasisfied, all two-setting Bell experiments in the
state $\rho$ have local 
realistic theories \cite{Zukowski1}.
Here, $\sigma_{1}$ and $\sigma_{2}$ are Pauli spin operators satisfying 
anti-commuting relation $\{\sigma_{1}, \sigma_{2}\}={\bf 0}$.
Especially, the condition (\ref{Z}) is a necessary and sufficient condition for the particular set of states (\ref{GHZ_WERNER}) for the existence of local realistic theories for all two-setting Bell experiments.
When $V\leq \frac{1}{2^{(N-1)/2}}$, all two-setting Bell experiments in the state $\rho_V$ are reproducible by
local realistic theories. 
In other words, all standard Bell inequalities are satisfied.
Assume $V= \frac{1}{2^{(N-1)/2}}$.
One has
\begin{equation}
|{\rm Tr}(B_N \rho_V)|=2^{(N-1)/2}V = 1.
\end{equation}
The positive partial transpose inequality (\ref{NPPT_BOUND}) is stronger 
than the above value by a factor of $2^{(N-1)/2}$.
Thus, we have shown that positivity of {\it every} partial transpose 
of $N$-qubit quantum states 
implies new inequalities on Bell correlations 
which are stronger than 
standard Bell inequalities by a factor of $2^{(N-1)/2}$.

In convenience, 
we introduce so-called `positive partial transpose operator' as
\begin{eqnarray}
P_N=2^{(N-1)}(|\psi^+ \rangle \langle \psi^+| 
- |\psi^- \rangle \langle \psi^-| ).
\end{eqnarray}
Then, the positive partial transpose inequlaity (\ref{NPPT_BOUND}) is 
expressed as
\begin{eqnarray}
|\langle P_N\rangle|\leq 1.\label{FINAL}
\end{eqnarray}

\section{Multipartite Bound entangled states}\label{BOUND}

In this section, we shall show that a family of D\"ur bound entangled states violates the positive partial transpose inequality 
(\ref{FINAL}) for $N\geq 4$.
Such a bound entangled state was introduced as follows
by D\"ur \cite{DUR}:
\begin{equation}
\rho_N = \frac{1}{N+1} 
\left( |\psi^+ \rangle \langle \psi^+| + \frac{1}{2} \sum_{k=1}^N (P_k + \tilde P_k) \right),
\end{equation}
with $P_k$ being 
a projector on the state 
$|0 \rangle_1 ... |1 \rangle_k ... |0 \rangle_N$ with ``1'' 
on the $k$th position 
($\tilde P_k$ is obtained from $P_k$ after replacing ``0'' by ``1'' and vice versa).
As originally shown in \cite{DUR}
this state violates the Bell-Mermin inequality for $N \ge 8$.
It is easy to see that.
The Bell-Mermin inequality $|\langle B_N \rangle |\leq 1$
predicts the violation factor of:
\begin{equation}
{\rm Tr}[B_N\rho_N] = \frac{2^{(N-1)/2}}{N+1},
\label{viol-bes}
\end{equation}
which comes from the contribution
of the GHZ state $| \psi^+ \rangle$
to the bound entangled state by D\"ur.
We find that $\frac{2^{(N-1)/2}}{N+1}>1$ when $N\geq 8$.
Thus, D\"ur bound entangled states 
violate the Bell-Mermin inequality for $N\geq 8$.
From the argument presented in Ref.~\cite{ACIN}, a family of D\"ur bound entangled states has a property 
that they are bipartite distillable entangled states for $N\geq 8$.

After that, it was shown that 
D\"ur bound entangled states violate Bell inequality \cite{ZUK_KASZ} with three settings per site for $N\geq 7$ \cite{KASZLI}.
Let us investigate the phenomenon.
The Bell operator of the Bell inequality with three settings per site
$|\langle B(3)_N \rangle |\leq 1$
is as \cite{NAGATA1}
\begin{eqnarray}
B(3)_N=\frac{1}{\sqrt{3}}(\frac{3}{2})^N
(|\psi^+ \rangle \langle \psi^+| 
- |\psi^- \rangle \langle \psi^-| ).
\end{eqnarray}
Hence, the Bell inequality
predicts the violation factor of:
\begin{equation}
{\rm Tr}[B(3)_N\rho_N] = \frac{1}{\sqrt{3}}(\frac{3}{2})^N\frac{1}{N+1},
\label{viol-bes}
\end{equation}
which comes from the contribution
of the GHZ state $| \psi^+ \rangle$
to the bound entangled state by D\"ur.
We find that $ \frac{1}{\sqrt{3}}(\frac{3}{2})^N\frac{1}{N+1}>1$ 
when $N\geq 7$.
Thus, D\"ur bound entangled states 
violate the Bell inequality for $N\geq 7$.
From similar to the argument presented in Ref.~\cite{ACIN}, a family of D\"ur bound entangled states has a property 
that they are bipartite distillable entangled states for $N\geq 7$.

It also turned out that D\"ur bound entangled states violate Bell inequality \cite{Zukowski0} with a continuous range of settings of the apparatus at each site for $N\geq 6$ \cite{ADITI}.
Let us investigate the phenomenon.
The Bell operator of the Bell inequality with a continuous range of settings of the apparatus at each site 
$|\langle B(\infty)_N \rangle |\leq 1$
is as \cite{NAGATA1,NAGATA3}
\begin{eqnarray}
B(\infty)_N=\frac{1}{2}(\frac{\pi}{2})^N
(|\psi^+ \rangle \langle \psi^+| 
- |\psi^- \rangle \langle \psi^-| ).
\end{eqnarray}
Hence, the Bell inequality
predicts the violation factor of:
\begin{equation}
{\rm Tr}[B(\infty)_N\rho_N] = \frac{1}{2}(\frac{\pi}{2})^N\frac{1}{N+1},
\label{viol-bes}
\end{equation}
which comes from the contribution
of the GHZ state $| \psi^+ \rangle$
to the bound entangled state by D\"ur.
We find that $ \frac{1}{2}(\frac{\pi}{2})^N\frac{1}{N+1}>1$ 
when $N\geq 6$.
We see D\"ur bound entangled states 
violate the Bell inequality for $N\geq 6$.
From similar to the argument presented in Ref.~\cite{ACIN}, a family of D\"ur bound entangled states has a property 
that they are bipartite distillable entangled states for $N\geq 6$.

The positive partial transpose inequality (\ref{FINAL})
predicts the substantially bigger violation factor of:
\begin{equation}
{\rm Tr}[P_N\rho_N] = \frac{2^{(N-1)}}{N+1},
\label{viol-bes}
\end{equation}
which comes from the contribution
of the GHZ state $| \psi^+ \rangle$
to the bound entangled state by D\"ur.
We find that $\frac{2^{(N-1)}}{N+1}>1$ when $N\geq 4$.
Thus, D\"ur bound entangled states 
violate the new inequality for $N\geq 4$.
From similar to the argument presented in Ref.~\cite{ACIN}, a family of D\"ur bound entangled states has a property 
that they are bipartite distillable entangled states for $N\geq 4$.


\section{summary}\label{summary}

In summary, we have shown that positivity of {\it every} partial transpose 
of $N$-qubit quantum states 
implies new inequalities on Bell correlations 
which are stronger than 
standard Bell inequalities by a factor of $2^{(N-1)/2}$.
A violation of the inequality 
implies that the system is in bipartite distillable entanglement 
since there is at least one bipartite splitting of the system such that 
the state becomes distillable.
It turned out that D\"ur bound entangled states 
violate the new inequality for $N\geq 4$.
Thus, a family of D\"ur bound entangled states has a property 
that they are bipartite distillable entangled states for $N\geq 4$.

\acknowledgments
This work has been
supported by Frontier Basic Research Programs at KAIST and K.N. is
supported by the BK21 research professorship.


\begin{thebibliography}{9}


\bibitem{BELL}
J. S. Bell, Physics (Long Island City, N.Y.) {\bf 1}, 195 (1964).


\bibitem{bib:Redhead}
M. Redhead,
{\it Incompleteness, Nonlocality, and Realism}, 
(Clarendon Press, Oxford, 1989), 2nd ed.

\bibitem{bib:Peres}
A. Peres, 
{\it Quantum Theory: Concepts and Methods}
(Kluwer Academic, Dordrecht, The Netherlands, 1993).





\bibitem{WERNER1}
R. F. Werner, 
Phys. Rev. A {\bf 40}, 4277 (1989).








\bibitem{DUR1}
W. D\"ur, J. I. Cirac, and R. Tarrach, 
Phys. Rev. Lett. {\bf 83}, 3562 (1999).

\bibitem{DUR2}
W. D\"ur and J. I. Cirac, 
Phys. Rev. A {\bf 61}, 042314 (2000).

\bibitem{KRAUS}
B. Kraus, J. I. Cirac, S. Karnas, and M. Lewenstein,
Phys. Rev. A {\bf 61}, 062302 (2000).


\bibitem{KARNAS}
S. Karnas and M. Lewenstein,
Phys. Rev. A {\bf 64}, 042313 (2001).




\bibitem{PPT}
A. Peres, 
Phys. Rev. Lett. {\bf 77}, 1413 (1996);
M. Horodecki, P. Horodecki, and R. Horodecki, 
Phys. Lett. A {\bf 223}, 1 (1996).

\bibitem{HORODECKI}
M. Horodecki, P. Horodecki, and R. Horodecki, 
Phys. Rev. Lett. {\bf 80}, 5239 (1998).








\bibitem{NAGATA1}
K. Nagata, W. Laskowski, and T. Paterek,
Phys. Rev. A {\bf 74}, 062109 (2006).


\bibitem{PERES}
A. Peres,
Found. Phys, {\bf 29}, 589 (1999).




\bibitem{Nielsen}
M. A. Nielsen and I. L. Chuang, {\it Quantum
Computation and Quantum Information} 
(Cambridge University Press, Cambridge, 2000).

\bibitem{Galindo}
A. Galind and M. A. Mart\'\i n-Delgado,
Rev. Mod. Phys. {\bf 74}, 347 (2002).



\bibitem{Bennett1}
C. H. Bennett, G. Brassard, S. Popescu, B. Schumacher, J. A. Smolin,
and W. K. Wootters,
Phys. Rev. Lett. {\bf 76}, 722 (1996).

\bibitem{Bennett2}
C. H. Bennett, H. J. Bernstein, S. Popescu, and B. Schumacher,
Phys. Rev. A {\bf 53}, 2046 (1996).





\bibitem{DUR}
W. D\"ur,
Phys. Rev. Lett. {\bf 87}, 230402 (2001).

\bibitem{Mermin}
N. D. Mermin, 
Phys. Rev. Lett. {\bf 65}, 1838 (1990);
S. M. Roy and V. Singh,
Phys. Rev. Lett. {\bf 67}, 2761 (1991);
A. V. Belinskii and D. N. Klyshko,
Phys. Usp. {\bf 36}, 653 (1993).


\bibitem{ACIN}
A. Ac\'\i n,
Phys. Rev. Lett. {\bf 88}, 027901 (2001).
cf. A. Ac\'\i n, V. Scarani, and M. M. Wolf,
Phys. Rev. A {\bf 66}, 042323 (2002) and the rich bibliography therein.


\bibitem{ZUK_KASZ}
M. \.Zukowski and D. Kaszlikowski, 
Phys. Rev. A {\bf 56}, R1682 (1997).


\bibitem{KASZLI}
D. Kaszlikowski, L. C. Kwek, J. Chen, and C. H. Oh,
Phys. Rev. A {\bf 66}, 52309 (2002).



\bibitem{Zukowski0}
M. \.Zukowski,
Phys. Lett. A {\bf 177}, 290 (1993).




\bibitem{ADITI}
A. Sen (De), U. Sen, and M. \.Zukowski,
Phys. Rev. A {\bf 66}, 62318 (2002).


\bibitem{ROY}
S. M. Roy, 
Phys. Rev. Lett. {\bf 94}, 010402 (2005).






\bibitem{Zukowski1}
M. \.Zukowski and \v{C}. Brukner, 
Phys. Rev. Lett. {\bf 88}, 210401 (2002).


\bibitem{Werner2}
R. F. Werner and M. M. Wolf,
Phys. Rev. A {\bf 64}, 032112 (2001);
R. F. Werner and M. M. Wolf,
Quant. Inf. Comp. {\bf 1}, 1 (2001).




\bibitem{Werner3}
R. F. Werner and M. M. Wolf, 
Phys. Rev. A {\bf 61}, 062102 (2000).




\bibitem{NAGATA2}
K. Nagata,
Phys. Rev. A {\bf 66}, 064101 (2002).


\bibitem{GHZ}
D. M. Greenberger, M. A. Horne, and A. Zeilinger,
in {\it Bell's Theorem, Quantum Theory and Conceptions of the Universe},
edited by M. Kafatos (Kluwer Academic, Dordrecht, The Netherlands, 
1989), pp. 69-72;
D. M. Greenberger, M. A. Horne, A. Shimony, and A. Zeilinger,
Am. J. Phys. {\bf 58}, 1131 (1990).



\bibitem{NAGATA3}
K. Nagata and J. Ahn,
arXiv:quant-ph/0302090.







































\end{thebibliography}
\end{document}